\documentclass[useAMS]{mnras}

\usepackage{graphicx}
\usepackage[fleqn]{amsmath}
\usepackage{multirow}
\usepackage{hyperref}
\usepackage{xspace}
\usepackage{url}

    \newcommand{\msun}{$M_{\hbox{$\odot$}}$\xspace}
    \newcommand{\lsun}{$L_{\hbox{$\odot$}}$\xspace}

    \newcommand{\kms}{\hbox{km s$^{-1}$}\xspace}

    \newcommand{\lumcgs}{\hbox{erg s$^{-1}$}\xspace}
    
    \newcommand{\fek}{Fe\,K\xspace}

    \newcommand{\hb}{H$\beta$\xspace}

    \newcommand{\nh}{N_{\rm H}}
    \newcommand{\mh}{m_{\rm p}}

    \newcommand{\lbol}{L_{\rm bol}}
    \newcommand{\lagn}{L_{\rm AGN}}

    \newcommand{\ledd}{L_{\rm Edd}}
    
    \newcommand{\mbh}{$M_{\rm BH}$}
    \newcommand{\rg}{r_{\rm g}}
    \newcommand{\rgdef}{GM_{\rm BH}/c^2}

    \newcommand{\nx}{N_{24}}

    \newcommand{\chandra}{\emph{Chandra}\xspace}
    \newcommand{\nustar}{\emph{NuSTAR}\xspace}

    \newcommand{\mrk}{Mrk\,231\xspace}
    \newcommand{\iras}{IRAS\,F11119+3257\xspace}

\title[AGN outflows and accretion history]{Multi-phase outflows as probes of AGN accretion history}
\author[E. Nardini and K. Zubovas]
{Emanuele Nardini$^1$\thanks{E-mail: enardini@arcetri.astro.it} and 
Kastytis Zubovas$^{2,3}$\\
$^1$INAF -- Osservatorio Astrofisico di Arcetri, Largo Enrico Fermi 5, I-50125 Firenze, Italy\\
$^2$Center for Physical Sciences and Technology, Saul{\.e}tekio av. 3, Vilnius LT-10257, Lithuania\\
$^3$Vilnius University Observatory, Saul{\.e}tekio av. 9, Bldg III, Vilnius LT-10222, Lithuania}

\begin{document}
%%%%%%%%%%%%%%%%%%%%%%%%%%%%%%%%%%%%%%%%%%%%%%%%%%%%%%%%%%%%%%%%%%%%%%%%%%%

\date{Released Xxxx Xxxxx XX}

\pagerange{\pageref{firstpage}--\pageref{lastpage}} \pubyear{2018}

\maketitle

\label{firstpage}

%%%%%%%%%%%%%%%%%%%%%%%%%%%%%%%%%%%%%%%%%%%%%%%%%%%%%%%%%%%%%%%%%%%%%%%%%%%
\begin{abstract}
Powerful outflows with a broad range of properties (such as velocity, ionization, radial 
scale and mass loss rate) represent a key feature of active galactic nuclei (AGN), even more 
so since they have been simultaneously revealed also in individual objects. Here we revisit 
in a simple analytical framework the recent remarkable cases of two ultraluminous infrared 
quasars, \iras and \mrk, which allow us to investigate the physical connection between 
multi-phase AGN outflows across the ladder of distance from the central supermassive black 
hole (SMBH). We argue that any major deviations from the standard outflow propagation 
models might encode unique information on the past SMBH accretion history, and briefly 
discuss how this could help address some controversial aspects of the current picture of 
AGN feedback.
\end{abstract}


%%%%%%%%%%%%%%%%%%%%%%%%%%%%%%%%%%%%%%%%%%%%%%%%%%%%%%%%%%%%%%%%%%%%%%%%%%%
\section{Introduction}

The past few years have witnessed a rapid progress in the detection and characterization 
of outflow signatures in active galactic nuclei (AGN), also thanks to the development of 
new powerful techniques based on interferometry and integral field spectroscopy. In the most 
luminous objects at any redshift, the inferred mass and momentum rates of the outflowing gas 
are much larger than the star formation rate of the host galaxy and the momentum rate of the 
AGN radiation field, respectively (e.g. Maiolino et al. 2012; Rupke \& Veilleux 2013; Liu 
et al. 2013; Cicone et al. 2014; Carniani et al. 2015; Bischetti et al. 2017; 
Gonz{\'a}lez-Alfonso et al. 2017). Outflows of this kind are therefore the best candidates 
for the widely invoked feedback agency that solves several open issues with galaxy evolution 
models (Kormendy \& Ho 2013, and references therein), bridging the small and the large scales 
by returning to the host galaxy a significant fraction of the energy released during the 
growth of the central supermassive black hole (SMBH). The actual impact of AGN-driven 
%outflows, however, is still unclear, as the observable consequences on the ability of the 
%host systems to sustain their star formation activity depend on a number of factors, such 
%as sample selection, adopted diagnostics, and temporal delays, thus delivering controversial 
%indications (see Harrison 2017 for a recent review, and references therein). 
outflows on the ability of the host systems to sustain their star formation activity, 
however, is still unclear, as the observable effects depend on a number of factors, 
such as sample selection, adopted diagnostics and temporal delays, thus delivering 
controversial indications (see Harrison 2017 for a recent review, and references therein).

Despite these uncertainties, the mounting evidence for galaxy-wide molecular outflows  
apparently complying with the adiabatic expansion of hot bubbles inflated by ultra-fast, 
accretion-disc winds (e.g. Tombesi et al. 2015) represents a potential breakthrough in many 
respects. The coexistence of multiple outflow flavours in individual objects is definitely 
intriguing, but at the same it poses some interpretational challenges. In fact, it is not 
obvious that the evolution of a SMBH wind can be tracked simultaneously at various stages 
in a single source, as some of these are arguably short-lived (at least in terms of 
detectability). If anything, the different response times to changes in the AGN activity 
level will reduce the probability that simultaneity also implies causal connection. 
The optimal conditions for the launch of high-velocity, high-column density X-ray 
winds, for instance, are met at the highest accretion rates (King \& Pounds 2003; 
Takeuchi, Ohsuga \& Mineshige 2013), close to or exceeding the Eddington limit, which 
cannot be preserved over periods comparable to the entire outflow lifetimes, which are 
typically of several million years. Moreover, while the emission signatures of neutral 
outflowing gas can still be visible at the largest scales a long time after the driving 
AGN episode has faded (King, Zubovas \& Power 2011), absorption and highly ionized features 
are unlikely to be detached from the parent activity spell. Numerical simulations (Gabor 
\& Bournaud 2014), as well as empirical (Schawinski et al. 2015) and theoretical arguments 
(King \& Nixon 2015), suggest that AGN have a flickering behaviour, with bursts as short 
as $\sim$\,10$^5$ yr. A slowly fading AGN luminosity after such Eddington-limited `flares' 
would still correlate with the outflow properties on kpc scales (Zubovas 2018), possibly 
alleviating but not completely solving the time-scale issues. 

In this paper, we revisit some of the most recent findings on multi-phase AGN outflows, 
discussing not only their implications in the context of the driving mechanism and 
propagation through the host galaxy, but also the potential application of outflows 
as a powerful means of probing the accretion history of the SMBH itself. Over the 
lifetime of a galaxy, a flickering AGN can generate an `outflow cascade' that will 
expand into an ever different medium, especially in chaotic accretion (Nayakshin, 
Power \& King 2012) and merger scenarios (Capelo et al. 2015). What we observe today 
might then be the compound of several previous events of highly efficient accretion. 
The prospect of disentangling the nuclear activity history by inspecting the properties 
of multi-phase outflows foreshadows novel insights into the way galaxies are shaped 
by their central engines. 

%%%%%%%%%%%%%%%%%%%%%%%%%%%%%%%%%%%%%%%%%%%%%%%%%%%%%%%%%%%%%%%%%%%%%%%%%%%
\section{Unification of AGN outflows}

Most AGN exhibit some kind of outflow signatures, which, depending on their exact nature, 
might arise over a wide range of distances from the nucleus. Mildly relativistic winds 
pervading the SMBH neighbourhood are identified through blueshifted absorption lines from 
highly ionized iron in the hard X-ray spectra (Tombesi et al. 2010), and are regarded as 
the ultimate trigger for AGN feedback (e.g. Nardini et al. 2015). Moving towards larger radii 
(and lower ionization states), the most prevalent outflow tracers materialize as soft X-ray 
and ultraviolet (UV) warm absorbers (Crenshaw \& Kraemer 2012), UV broad absorption lines 
(BALs; Gibson et al. 2009), blue wings in the optical absorption (Rupke \& Veilleux 2011) 
and emission lines (Harrison et al. 2014), molecular P-Cygni profiles in the far-infrared 
(Sturm et al. 2011) and broad, spatially resolved emission lines in the sub-millimetric 
(Feruglio et al. 2010). In some of these cases, the typical ranges of both distance and 
velocity partly overlap. For the sake of simplicity, throughout this paper we use the terms 
`wind' and `outflow' to describe, respectively, the faster/inner phases and the slower/outer 
ones. 

Whether these disparate outflow manifestations, often characteristic of AGN with different 
luminosities, redshifts and host-galaxy properties, can be considered as consecutive snapshots 
of the same process is far from straightforward. At least in the quasar luminosity regime 
($\lbol \sim 10^{46}~\lumcgs$), the compilation of X-ray winds and molecular outflows from 
the literature is compatible with the predictions of inefficient cooling of the shocked 
gas (Zubovas \& King 2012; Faucher-Gigu{\`e}re \& Quataert 2012). Yet, uncertainties and 
systematics are extremely large, and still not compensated for by sufficient sample statistics. 
The recent study by Fiore et al. (2017) has confirmed a strong correlation between the properties 
of each outflow flavour (including also [O\,\textsc{iii}], BALs and warm absorbers) and AGN 
luminosity, but the connection between the various phases remains unclear (see their Fig. 2). 
The key physical parameters of two coexisting X-ray and molecular components were first measured 
by Tombesi et al. (2015), who discovered strongly blueshifted \fek absorption in \iras, an 
ultraluminous infrared galaxy (ULIRG) already known to host a massive OH outflow (Veilleux et 
al. 2013b). The relation between \fek and OH energetics was suggested to support the scenario 
of an energy-conserving flow. This case is re-examined below, together with another remarkable 
example that became available soon afterwards, \mrk  (Feruglio et al. 2015). 

\subsection{Mass outflow rates}
In order to maximize the information that can be extracted from the comparison of different 
phases, we need to accurately assess the amount of energy carried outwards by the accretion-disc 
wind in the first place. Here we compute the initial mass outflow rate through the expression: 
\begin{equation}
\dot{\mathcal{M}}_{\rm wind} \simeq \Omega \nh \mh \mathcal{V} \mathcal{R},
\label{e0}	
\end{equation}
implicitly assuming solar abundances, full ionization, and neglecting any more complex 
geometrical dependence with no significant loss of accuracy (e.g. Krongold et al. 2007). 
In Eq.\,(\ref{e0}), $\mh$ is the proton mass, $\Omega$ is the solid 
angle subtended by the wind, and $\nh$, $\mathcal{V}$ and $\mathcal{R}$ are its column 
density, velocity and inner radius. As, with very few exceptions, solid angle and starting 
point cannot be properly constrained from the X-ray spectrum, we adopt $\Omega_{\rm FeK}/4\upi 
= 0.5$ (Nardini et al. 2015) and $\mathcal{R}_{\rm FeK} = 2c^2/\mathcal{V}_{\rm FeK}^2$ 
(i.e., the escape radius in units of gravitational radii, $\rg = \rgdef$), so that: 
\begin{equation}
\dot{\mathcal{M}}_{\rm FeK} \simeq 9.4 \, \nx M_8 \beta^{-1} \times 10^{24}~\rm{g~s}^{-1}, 
\label{e1}
\end{equation}
where $\nx$ is the column density in units of 10$^{24}$ cm$^{-2}$, $M_8$ is the SMBH mass 
in units of 10$^8$\,\msun, and $\beta = \mathcal{V}_{\rm FeK}/c$. This is roughly equivalent 
to 0.15\,$\nx M_8 \beta^{-1}$ \msun yr$^{-1}$.

The mass outflow rate of the large-scale components is instead averaged over their flow 
time, $\dot{\mathcal{M}}_{\rm out} = \mathcal{M} \mathcal{V} \mathcal{R}^{-1}$, as this is 
more appropriate for the comparison with a global outflow dynamics powered by a single 
AGN episode of constant luminosity. Hence, for the molecular gas phases, it is: 
\begin{equation}
\dot{\mathcal{M}}_{\rm out} \simeq 6.5 \, \mathcal{M}_8 \mathcal{V}_3 \mathcal{R}_{\rm kpc}^{-1} 
\times 10^{27}~\rm{g~s}^{-1},
\label{e2}
\end{equation}
where $\mathcal{M}_8$ is the total outflowing mass in 10$^8$\,\msun, $\mathcal{V}_3$ is 
the velocity in 10$^3$ km s$^{-1}$, $\mathcal{R}_{\rm kpc}$ is the distance in kpc, and 
the numerical factor is about 100 \msun yr$^{-1}$. A `fiducial' radius is frequently used 
in Eq.\,(\ref{e2}), yet here we consistently identify $\mathcal{R}_{\rm kpc}$ with the 
smallest spatial scale affected by the outflow. The meaning of $\mathcal{R}$ in 
Eqs.\,(\ref{e0}--\ref{e2}) thus implies that our estimates of $\dot{\mathcal{M}}$ and 
of its dependent quantities should be treated as lower/upper limits for the wind/outflow 
cases. The motivations behind this conservative choice will become clearer in Section~3, 
once all the relevant pieces of information on the sources under examination are in hand.

\subsection{Black hole masses}
To begin with, it is immediately evident from Eq.\,(\ref{e1}) that a reliable estimate of 
the SMBH mass is very important. This also controls another crucial ingredient of our 
analysis, the Eddington ratio: the lower the mass, the higher the (average) accretion 
rate the AGN has to maintain to drive an outflow towards the edge of the galaxy. The 
values of $M_8$ reported in the literature for both \iras and \mrk span more than an order 
of magnitude. We therefore resort to the latest scaling relations between virial radius of 
the broad line region and black hole mass (Bentz et al. 2013), and apply them to the 
de-reddened optical spectra of Zheng et al. (2002; see their discussion of the correction 
for extinction local to the source). For a full width at half-maximum (FWHM) of the \hb 
line of $\sim$\,2000 \kms, intrinsic $\lambda L_\lambda (5100$\,\AA) $\simeq 12.1$\,\lsun, 
and virial coefficient $f = 1$, we evaluate a SMBH mass in \iras of 
$M_8 = 1.95^{+0.80}_{-0.54}$ (see also Hao et al. 2005). Note that the most recent works 
on this object have used an appreciably smaller value ($M_8 = 0.16$, after Kawakatu, 
Imanishi \& Nagao 2007), which, however, turns out to be underestimated by a factor of 
$\sim$\,5 (N. Kawakatu, private communication). 

For \mrk, exhibiting a FWHM\,(\hb) $\simeq 3100$ \kms and an extinction-corrected 
$\lambda L_\lambda (5100$\,\AA) $\simeq 11.5$\,\lsun, we find $M_8 = 2.40^{+0.55}_{-0.40}$. 
This is remarkably similar to the mass inferred by Leighly et al. (2014) employing the 
Pa$\alpha$ line width and the 1\,$\mu$m continuum luminosity (following Landt et al. 
2013). Considerably lower and fairly higher values are returned, respectively, by the 
correlations with the stellar velocity dispersion from CO 1.6\,$\mu$m ro-vibrational 
band heads ($M_8 = 0.17$; Dasyra et al. 2006) and with the $H$-band magnitude of the 
host spheroid ($M_8 = 3.8$; Veilleux et al. 2009a). Yet, the former method is known 
to yield systematically small velocity dispersions (e.g. Rothberg et al. 2013), while 
the latter, in these hybrid, IR-bright systems, might suffer from heavy starburst 
contamination. Consequently, the corresponding dynamical and photometric black hole 
masses are most likely extremes, and are deemed to be less reliable than the virial 
ones we have retrieved above. 

%\vspace*{3pt}
%\noindent
%\textbf{\iras.}
%\textbf{\underline{\smash{\iras.}}} 
%$\bullet$ 
\subsection{\iras}
As previously mentioned, the unexpected detection of an ultra-fast wind in \iras has 
opened unprecedented perspectives. Since minor variations between different epochs are 
negligible for our purposes, the gas velocity, $\beta = 0.253^{+0.061}_{-0.118}$, and 
column density, $\nx = 3.2\,(\pm 1.5)$, are derived from a follow-up observation of the 
source performed with \nustar (Tombesi et al. 2017), which samples the continuum also 
beyond the \fek absorption feature and so provides tighter constraints on its position 
($\beta$) and depth ($\nx$). By using these values and $M_8 = 1.95$ in Eq.\,(\ref{e1}), 
we obtain mass and momentum rates for the X-ray wind of 
$\dot{\mathcal{M}}_{\rm FeK} \sim 1.5$--6.5 \msun yr$^{-1}$ and 
$\dot{\mathcal{P}}_{\rm FeK} = \dot{\mathcal{M}}_{\rm FeK} \mathcal{V}_{\rm FeK} 
\sim 0.8$--$2.8 \times 10^{36}$ dyne. 
%and $\dot{E}_{\rm kin,X} \sim 2$--$11 \times 10^{45}$ \lumcgs. 
It is worth noting that the latter, under our prescription for $\mathcal{R}_{\rm FeK}$, 
does not explicitly depend on $\beta$.

The properties of the OH 119\,$\mu$m outflow were determined by Tombesi et al. (2015) 
through the radiative transfer model described in Gonz{\'a}lez-Alfonso et al. (2014), 
for a relative OH abundance of $2.5 \times 10^{-6}$. The absorbing material was constrained to 
lie within 0.1--1.0 kpc from the SMBH, with $\mathcal{V}_{\rm OH} \simeq 1000\,(\pm 200)$ 
\kms. Based on our assumptions, we get $\dot{\mathcal{M}}_{\rm OH} \sim 200$--1500 \msun 
yr$^{-1}$ and $\dot{\mathcal{P}}_{\rm OH} \sim 1$--$9 \times 10^{36}$ dyne. 
%and $\dot{E}_{\rm kin,OH} \sim 1.9 \times 10^{44}$ \lumcgs. 
Subsequent ALMA observations have also revealed broad wings (with nominal velocity 
$\mathcal{V}_{\rm CO} = 1000$ \kms) in the CO\,(1--0) emission line profile, spatially 
extended over distances of $\sim$\,4--15 kpc (Veilleux et al. 2017). The standard 
conversion factor of 0.8 between CO line luminosity and total molecular (H$_2$) mass 
in ULIRGs (Downes \& Solomon 1998) delivers $\mathcal{M}_8 \sim 10\,(\pm 4)$, hence 
%\footnote{This is $\alpha_{\rm CO} = 0.8$ \msun (K \kms pc$^2$)$^{-1}$ (Downes \& Solomon 1998), and it is about 5 times lower than in the Milky Way. In this work, we use a single value of $\alpha_{\rm CO}$ irrespective of the specific transition.} 
$\dot{\mathcal{M}}_{\rm CO} \sim 140$--370 \msun yr$^{-1}$ and 
$\dot{\mathcal{P}}_{\rm CO} \sim 0.7$--$2.5 \times 10^{36}$ dyne.
%and $\dot{E}_{\rm kin,CO} \sim 8.1 \times 10^{43}$ \lumcgs. 
The values of all the quantities of interest are summarized in Table~\ref{st}.

The three main outflow flavours in \iras, \fek, OH and CO, are reported as red points in the 
momentum rate versus velocity diagram of Fig.~\ref{wd}, with no further corrections. Notably, 
evidence for widespread outflows in this source is also found in other gas phases, in 
particular Na\,\textsc{i}\,D (Rupke, Veilleux \& Sanders 2005), 
[O\,\textsc{iii}] (L{\'{\i}}pari 
et al. 2003) and [Ne\,\textsc{v}] (Spoon \& Holt 2009), but unfortunately none of these has 
been investigated as yet in sufficient detail to be assigned a precise location in this plot.

\begin{table}
\centering
\small
\caption{AGN and outflow properties in \iras and \mrk. $\dot{M}_{\rm A}$ is the mass accretion rate, $\alpha$ the relative AGN contribution to the bolometric luminosity ($\lbol$), and $\eta$ the SMBH radiative efficiency. Note that the values provided for each outflow rate (mass, momentum, energy) neglect systematic uncertainties, and are meant to represent either lower (Fe K, BAL) or upper (OH, CO) limits (see the text for the full set of assumptions). OH entries for \mrk refer to the high-velocity components from Gonz{\'a}lez-Alfonso et al. (2017), corrected to match our criteria.}
\label{st}
\begin{tabular}{l@{\hspace{20pt}}c@{\hspace{20pt}}c}
\hline
 & IR\,11119 & \mrk \\
\hline
log\,(\mbh/\msun) & $8.29 \pm 0.15$ & $8.38 \pm 0.09$ \\[0.6ex]
%$M_{\rm BH}$ (10$^8$ \msun) & 1.95$^{+0.80}_{-0.54}$ & 2.40$^{+0.55}_{-0.40}$ \\[0.5ex]
log\,($\lbol$/\lsun) & 12.67 & 12.60 \\[0.6ex]
%$\lbol$ (10$^{12}$ \lsun) & & \\[1.2ex]
$\lagn/\ledd$ & $0.73 \alpha$ & $0.50 \alpha$ \\[0.6ex]
$\dot{M}_{\rm A}$ (\msun yr$^{-1}$) & $0.31 \alpha \eta^{-1}$ & $0.27 \alpha \eta^{-1}$ \\[1.5ex]
$\dot{\mathcal{M}}_{\rm FeK}$ (\msun yr$^{-1}$) & 3.6$^{+2.9}_{-2.1}$ & 1.4$^{+2.0}_{-0.8}$ \\[0.6ex]
$\dot{\mathcal{P}}_{\rm FeK}$ ($\lbol/c$) & 2.9$^{+1.9}_{-1.6}$ & 0.36$^{+0.48}_{-0.21}$ \\[0.6ex]
$\dot{\mathcal{E}}_{\rm FeK}$ ($\lbol$) & 0.37$^{+0.25}_{-0.27}$ & 0.012$^{+0.016}_{-0.007}$ \\[1.5ex]
$\dot{\mathcal{M}}_{\rm BAL}$ (\msun yr$^{-1}$) & $-$ & 12$^{+8}_{-4}$ \\[0.6ex]
$\dot{\mathcal{P}}_{\rm BAL}$ ($\lbol/c$) & $-$ & 0.67$^{+0.50}_{-0.32}$ \\[0.6ex]
$\dot{\mathcal{E}}_{\rm BAL}$ ($\lbol$) & $-$ & 0.005$^{+0.005}_{-0.003}$ \\[1.5ex]
$\dot{\mathcal{M}}_{\rm OH}$ (\msun yr$^{-1}$) & 600$^{+900}_{-400}$ & (1370) \\[0.6ex]
$\dot{\mathcal{P}}_{\rm OH}$ ($\lbol/c$) & 6.4$^{+9.6}_{-4.5}$ & (11.0) \\[0.6ex]
$\dot{\mathcal{E}}_{\rm OH}$ ($\lbol$) & 0.011$^{+0.017}_{-0.009}$ & (0.012) \\[1.5ex]
$\dot{\mathcal{M}}_{\rm CO}$ (\msun yr$^{-1}$) & $255 \pm 115$ & $870 \pm 330$ \\[0.6ex]
$\dot{\mathcal{P}}_{\rm CO}$ ($\lbol/c$) & $2.7 \pm 1.5$ & $9.2 \pm 4.5$ \\[0.6ex]
$\dot{\mathcal{E}}_{\rm CO}$ ($\lbol$) & 0.005$^{+0.003}_{-0.004}$ & $0.013 \pm 0.008$ \\
\hline
\end{tabular}
\end{table}

%\vspace*{3pt}
%\noindent
%\textbf{\mrk.} 
%$\bullet$ 
\subsection{\mrk}
Being an AGN-dominated ULIRG, \mrk qualifies as the nearest quasar known. Given its proximity 
and excellent observational coverage, it is the ideal target to explore all the possible 
connections between AGN-driven outflows. Indeed, \mrk displays almost every kind 
of outflow features commonly found among AGN (e.g. L{\'{\i}}pari et al. 2009; 
Feruglio et al. 2010; Fischer et al. 2010; Rupke \& Veilleux 2011; Gonz{\'a}lez-Alfonso 
et al. 2014; Aalto et al. 2015; Morganti et al. 2016). A very deep \chandra exposure in 
the X-rays has recently brought out also a moderately fast \fek wind ($\beta = 
0.067^{+0.007}_{-0.010}$; Feruglio et al. 2015), whose properties are determined 
as above. With $\nx = 0.27^{+0.36}_{-0.15}$ and $M_8 = 2.4$, from Eq.\,(\ref{e1}) we now obtain $\dot{\mathcal{M}}_{\rm FeK} \sim 0.6$--3.4 \msun yr$^{-1}$ 
and $\dot{\mathcal{P}}_{\rm FeK} \sim 0.7$--$4.3 \times 10^{35}$ dyne (Table~\ref{st}). 
%and $\dot{E}_{\rm kin,X} \simeq 2.3 \times 10^{44}$ \lumcgs. 

\mrk also belongs to the subclass of FeLoBALs, in that it shows broad UV absorption in 
several transitions of low-ionization species, including Fe\,\textsc{ii}. This allows 
us to probe an intermediate scale (in $\log \mathcal{R}$) between the accretion disc and 
the host galaxy. The location of the gas responsible for these features was estimated 
to be of the order of 0.1 kpc by Leighly et al. (2014), by applying a photoionization 
model, while Veilleux et al. (2016) argued for distances as small as a few pc, based on 
the physical conditions necessary to detect absorption from excited states of Fe\,\textsc{ii}. 
The BAL mass ouflow rate is then calculated using Eq.\,(\ref{e0}), assuming a covering 
factor of $\Omega_{\rm BAL}/4\upi = 0.2$ (Dunn et al. 2010) and a radius of 
$\mathcal{R}_{\rm BAL} = 2$ pc. Accordingly, this should be taken as a conservative 
measure as well. For $\mathcal{V}_{\rm BAL} \sim 4500\,(\pm 1000)$ \kms and 
$\nx \simeq 0.05^{+0.03}_{-0.01}$ (Leighly et al. 2014), 
we derive $\dot{\mathcal{M}}_{\rm BAL} \sim 8$--20 \msun yr$^{-1}$ and 
$\dot{\mathcal{P}}_{\rm BAL} \sim 1.8$--$6 \times 10^{35}$ dyne.
%and $\dot{E}_{\rm kin,BAL} \simeq 7.7 \times 10^{43}$ \lumcgs.

\begin{figure}
\includegraphics[width=8.5cm]{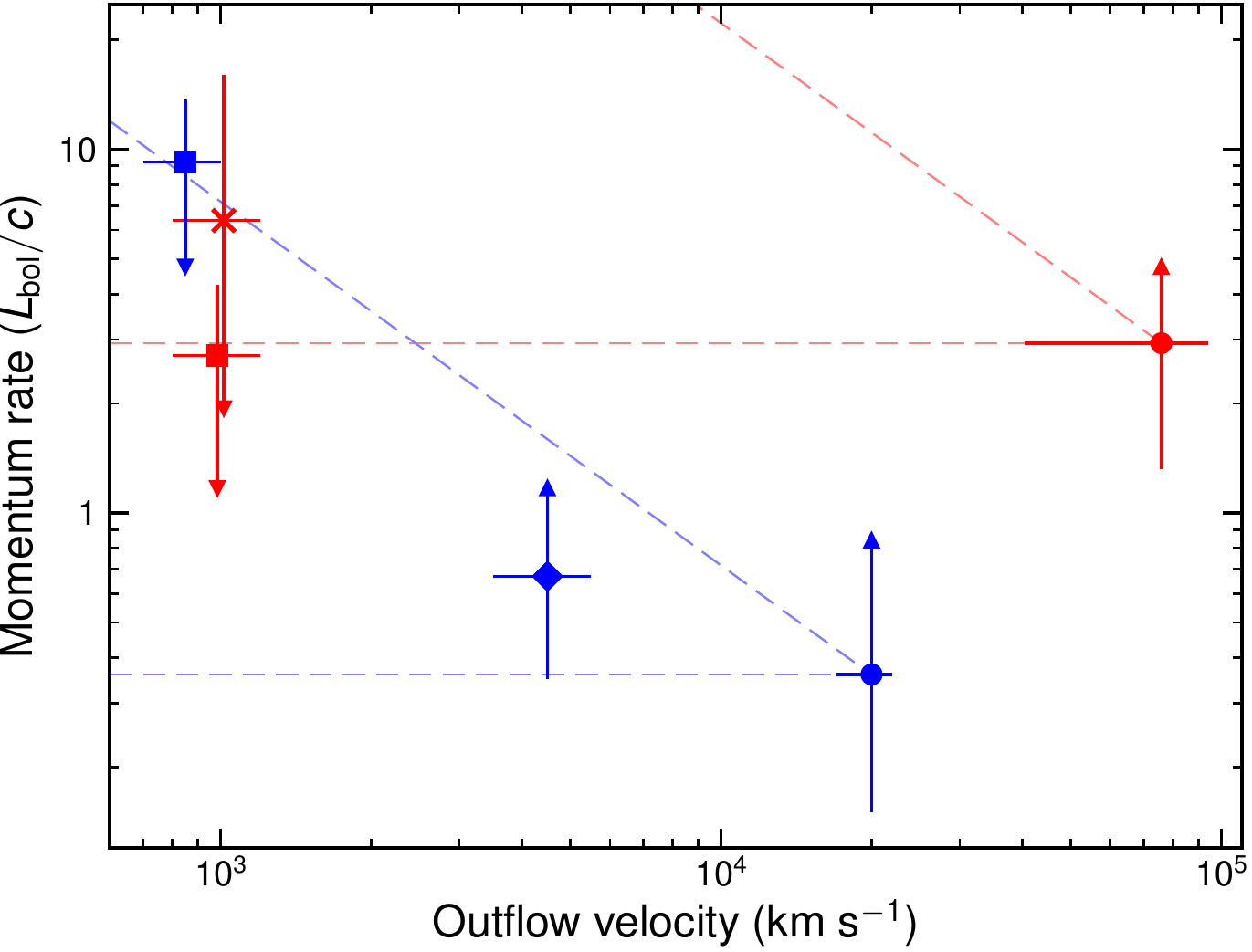}
%\vspace*{-10pt}
\caption{Momentum rate (in units of $\lbol/c$) against velocity of the multi-phase outflows 
in \iras (red) and \mrk (blue). The different symbols refer to the outflow type: \fek (dot), 
BAL (diamond), OH (cross), and CO (square). The dashed horizontal and diagonal lines 
correspond to momentum and energy conservation. Error bars (1$\sigma$) are purely 
statistical, while arrows indicate whether a given measure should be considered as 
a lower or an upper limit (see the text for details).} 
\label{wd}
\end{figure}

We finally consider the CO\,(2--1) outflow from Feruglio et al. (2015). The mass outflow 
rate has some mild radial dependence out to $\sim$\,1.3 kpc, with nearly constant 
velocity, 
$\mathcal{V}_{\rm CO} \simeq 850\,(\pm 150)$ \kms. Here we take into account the total 
mass, $\mathcal{M}_8 \simeq 3\,(\pm 1)$, which has been corrected for the slightly larger 
CO-to-H$_2$ conversion factor used in this work. For $\mathcal{R}_{\rm CO}$ = 0.3 
kpc, the resulting mass and momentum rates are, respectively, 
$\dot{\mathcal{M}}_{\rm CO} \sim 540$--1200 \msun yr$^{-1}$ and 
$\dot{\mathcal{P}}_{\rm CO} \sim 2.4$--$7 \times 10^{36}$ dyne. 
Feruglio et al. (2015) also showed that the neutral gas phase occupies a well-defined region 
of the momentum rate versus velocity diagram (Fig.~\ref{wd}), even when studied through alternative tracers. 
We can therefore take advantage of the wealth of available data on \mrk to independently 
support the inferred outflow rates. By converting to our definitions the parameters of the 
two high-velocity ($\mathcal{V}_{\rm OH} = 550$ and 700 \kms) OH components identified by 
Gonz{\'a}lez-Alfonso et al. (2017), we indeed achieve fully consistent reference values, 
which are listed within brackets in Table~\ref{st}. In the wake of this excellent match, 
in the remainder of this paper we can safely adopt our CO-based estimates as representative 
of the large-scale, neutral outflow.

%%%%%%%%%%%%%%%%%%%%%%%%%%%%%%%%%%%%%%%%%%%%%%%%%%%%%%%%%%%%%%%%%%%%%%%%%%%
\section{Discussion}

The diagram in Fig.~\ref{wd} has been mostly read so far as a diagnostic of the 
launching and driving mechanisms of AGN outflows. For this reason, the momentum rate is 
conveniently normalized to $\lagn/c$, giving the boost factor ($\mathcal{B}$) over the 
AGN radiative input. For a wind initially driven by radiation pressure in the Eddington 
regime, and then expanding adiabatically, the boost factor grows from $\mathcal{B} \sim 1$ 
at launch to $\mathcal{B} \sim 10$--20 at kpc scales (Zubovas \& King 2012). As ULIRGs 
are also host to fierce star formation, whose contribution to the total energy output 
is not trivial to disentangle from that of the AGN (e.g. Nardini et al. 2010), 
the measure of $\lagn$ is somewhat 
dependent on the adopted method. We thus prefer to normalize the momentum rates to 
$\lbol/c$, where the bolometric luminosity is $\lbol \sim 1.8$ and $1.5 \times 10^{46}$ 
\lumcgs for \iras and \mrk, respectively (Veilleux et al. 2009b). This only introduces 
a common, minor (both sources are widely believed to be AGN-dominated) shift downwards 
of each point in the `momentum boost' diagram of Fig.~\ref{wd}, without affecting their 
relative positions. Much larger uncertainties are associated to all the assumptions 
underlying our estimates, as emphasized earlier in our derivation of the mass (hence 
momentum) outflow rates. We refer to Harrison et al. (2018) for a more extensive 
discussion on these issues. Here we have resolved to partly keep the impact of 
systematics under control by not dealing with the `actual' values of $\dot{\mathcal{P}}$, 
but with indicative lower (\fek, BAL) and upper (OH, CO) limits.

In this context, we follow a simple analytical approach for the interpretation of 
Fig.~\ref{wd}, as this can be already very informative. At accretion rates close to 
Eddington, the wind's optical depth to electron scattering ($\tau$) is naturally 
around unity (King \& Pounds 2003; Reynolds 2012). Consequently, in the single-scattering 
limit, it is:
\begin{equation}
\dot{\mathcal{P}} = \dot{\mathcal{M}} \mathcal{V} \simeq \tau \frac{\lagn}{c} 
\sim \frac{\lagn}{c}.
\label{e3}
\end{equation}
If we now define $\dot{m}_{\rm A}$ and $\dot{m}_{\rm L}$ as the mass accretion and 
loss rates in units of Eddington, and $\eta$ as the SMBH radiative efficiency, two 
immediate ramifications of Eq.\,(\ref{e3}) are:
\begin{equation}
\eta \simeq \frac{\dot{m}_{\rm L}}{\dot{m}_{\rm A}} \frac{\mathcal{V}}{c} = 
\frac{\dot{m}_{\rm L}}{\dot{m}_{\rm A}} \beta 
\label{e4}
\end{equation}
and 
\begin{equation}
\dot{\mathcal{E}} = \frac{1}{2}\dot{\mathcal{M}} \mathcal{V}^2 \simeq \frac{1}{2}\beta \lagn,
\label{e5}
\end{equation}
where $\dot{\mathcal{E}}$ is the rate at which mechanical energy is injected into the host 
environment through the fast X-ray wind. These approximations can be reasonably applied to 
both \iras and \mrk, for which $\dot{m}_{\rm A} \sim 0.7 \alpha$ and $0.5 \alpha$, 
respectively, where $\alpha$ is the fractional AGN contribution to $\lbol$. Incidentally, 
we note that Eq.\,(\ref{e4}), with accurate data over large samples, could provide independent 
constraints on the SMBH spin distribution. Irrespective of the exact value of $\eta$, the 
range of $\dot{\mathcal{M}}_{\rm FeK}$ plainly encompasses also the mass accretion rate for 
both AGN (Table~\ref{st}). Given the sizeable uncertainties, in the following we can assume 
$\dot{m}_{\rm L} \sim \dot{m}_{\rm A}$ with no loss of generality, so that $\eta = \beta$. 

From Fig.~\ref{wd}, a striking contrast emerges between the various outflow manifestations 
in \iras (red) and in \mrk (blue). In the former source, despite a difference in distance 
of at least an order of magnitude, the OH and CO components have nearly identical velocity 
and very similar mass outflow rates, as if the outflow were not yet coasting but mass loading 
had stopped. Furthermore, both molecular phases lie rather far from the predictions for a fully 
energy-conserving case, based on the contemporary thrust of the \fek wind. This, in particular, 
is apparently at odds with the claim of Tombesi et al. (2015), who used a finite thickness of 
the OH shell to compute an `instantaneous' value of $\dot{\mathcal{P}}_{\rm OH}$, and corrected 
$\dot{\mathcal{P}}_{\rm FeK}$ downwards to match the inferred OH covering factor of 
$\sim$\,0.2. The 
properties in Fig.~\ref{wd} would entail a stable energy input from the AGN over several Myr 
(see also Veilleux et al. 2017), but a fairly small coupling efficiency between the fast and 
the slow components. Alternatively, $\dot{\mathcal{P}}_{\rm FeK}$ must be either overestimated 
or higher than in past epochs. Yet, our revised measure of the SMBH mass does not require the 
SMBH to accrete at super-Eddington rates, so the present activity could have been maintained 
for a period commensurate with the age of the CO outflow 
($\mathcal{R}_{\rm CO}/\mathcal{V}_{\rm CO} > 4$ Myr), especially in 
a ULIRG, where the gas supply to the central regions is governed by enhanced gravitational 
disturbances. 

This notwithstanding, it should be kept in mind that in our analysis we are comparing 
truly `instantaneous' (X-ray wind) and `time-averaged' (molecular outflow) quantities, so the 
ratio $\mathcal{T}_{\rm FeK}^{\rm CO} = \mathcal{E}_{\rm CO}/\dot{\mathcal{E}}_{\rm FeK}$ 
represents 
an equally effective but less biased indicator to constrain the wind efficiency. The 
mechanical energy of the CO phase integrated over its flow time (which only depends on 
observables: total outflowing mass and velocity) is 
$\mathcal{E}_{\rm CO} \sim 0.4$--$1.6 \times 10^{58}$ erg, and it can be provided by the 
SMBH wind in just $\mathcal{T}_{\rm FeK}^{\rm CO} \sim 0.01$--0.27 Myr. This range can be stretched 
up to $0.4\alpha^{-1}$ Myr by directly using Eq.\,(\ref{e5}), but this would neglect the 
fact that $\dot{\mathcal{P}}_{\rm FeK}$ in \iras formally exceeds $\lagn/c$ by a factor of 
$\sim$\,$3\alpha^{-1}$, consistent with the estimated optical depth of the wind, $\tau \geq 2.5$ 
($\nh$ is insensitive to any fully ionized gas). We thus conclude that 
$\mathcal{T}_{\rm FeK}^{\rm CO} \ll \mathcal{R}_{\rm CO}/\mathcal{V}_{\rm CO}$. Whether 
this is symptomatic of a real 
discrepancy between the energetics and lifetime of the CO outflow and the current AGN 
activity level is contingent upon the fraction of energy injected into the system through 
the fast wind that is eventually converted into bulk motion of the shocked ambient gas. 
This amount also depends on the density and metallicity of the interstellar medium (ISM), 
but it is typically of the order of $\sim$\,10--20 per cent of the input mechanical energy 
(e.g. Richings \& Faucher-Gigu{\`e}re 2017). The range for $\mathcal{T}_{\rm FeK}^{\rm CO}$ 
above is therefore $\sim$\,5--10 times shorter than the intrinsic wind lifetime, 
$\mathcal{T}_{\rm wind}$. None the less, 
some tension remains, which can be further mitigated by allowing for a combination 
of narrower 
opening angle and iron overabundance (hence smaller equivalent $\nh$), but is unlikely 
to be completely removed unless the source has also been caught in an unusual \fek outburst 
state, implying a high degree of fine-tuning. 

In this light, we cannot rule out the possibility that the CO outflow is actually the result 
of a previous Eddington-limited accretion episode. We follow this conjecture in the simplified 
analytical framework that describes the galactic environment as an isothermal sphere (e.g. 
Zubovas \& King 2012), according to which the relation between mass outflow rate and 
velocity of the swept-up gas is given by: 
\begin{equation}
\dot{\mathcal{M}} \simeq \frac{2f_{\rm g}\sigma^2}{G} \mathcal{V},
\label{e6}
\end{equation}
where $f_{\rm g}$ is the gas fraction and $\sigma$ is the velocity dispersion. The very 
presence of a large-scale outflow suggests that the SMBH in \iras has already reached 
the $M_{\rm BH}$--$\sigma$ relation (King 2010), so we can assume $\sigma = 180$ \kms (as 
appropriate for $M_8 \sim 2$; Kormendy \& Ho 2013) and apply Eq.\,(\ref{e6}) to the CO phase, 
deriving $f_{\rm CO} \sim 0.01$--0.025, which is much smaller than the cosmological value, 
$f_{\rm c} \simeq 0.16$. Were it still pushed by a constant driving force $\lagn/c$, such 
a tenuous gas component should travel at significantly larger velocity (King et al. 2011): 
\begin{equation}
\mathcal{V} \simeq \frac{\gamma + 1}{2}\left[\frac{2\eta \dot{m}_{\rm A} f_{\rm c}}{3 f_{\rm g}} 
\sigma^2c \right]^{1/3} \approx 3000~\kms,
\label{e7}
\end{equation}
with $\gamma$ specific heat ratio, $f_{\rm g} = f_{\rm CO}$ and $\eta \dot{m}_{\rm A} 
\sim \beta$ (Eq.~\ref{e4}). From this, as per equation (7) in Zubovas \& King (2012), we 
can infer the radius ($R_{\rm off}$) where the putative coasting stage started, after a 
sudden AGN switch-off:
\begin{equation}
\left(\frac{\mathcal{V}_{\rm CO}}{V_{\rm off}}\right)^2 \sim 3 \left(\frac{R_{\rm off}}{\mathcal{R}_{\rm CO}}\right)^2 
- 2 \left(\frac{R_{\rm off}}{\mathcal{R}_{\rm CO}}\right)^3.
\label{e7}
\end{equation}
In this scenario, the outflow must have lost radiative support when its distance 
from the AGN was about one fifth of the current one, which, for constant deceleration, 
corresponds to at least 1.5 Myr ago. This guess strongly depends on the actual radius 
of the CO shell, whose thickness of $\sim$\,10 kpc is indeed quite challenging, but 
widely separated AGN episodes are more likely to have distinct energy budgets (e.g. 
$\dot{\mathcal{E}}_{\rm FeK}$). We could argue that what we see on the large scales, 
although consistent with the steady driving from an AGN of lower, relatively constant 
luminosity (proportional to the duty cycle of Eddington-limited accretion; cf. Veilleux 
et al. 2017), is in fact the result of several 
short $\dot{m}_{\rm A} \sim 1$ bursts (e.g. Zubovas \& King 2016), only entraining the 
diffuse ISM of the galaxy while leaving behind the denser clumps. Molecules can form 
at later times within the outflow (Zubovas \& King 2014), as radiative cooling occurs 
within $\sim$\,1 Myr under most circumstances (Richings \& Faucher-Gigu{\`e}re 2017). 
Projection effects, however, are certainly involved, and radial profiles of velocity 
and mass outflow rate would be needed to distinguish between different histories. 
It is also worth noting that, if multiple gas phases coexist at the same radius, the 
total mass outflow rate can be easily underestimated, and so the required duration of 
Eddington-limited AGN activity ($\sim$\,$\mathcal{T}_{\rm wind}$). 

Turning now to the interpretation of the OH outflow in \iras, this is exempt from 
the main shortcomings that affect its CO counterpart. The SMBH wind can take up to 
$\mathcal{T}_{\rm FeK}^{\rm OH} = \mathcal{E}_{\rm OH}/\dot{\mathcal{E}}_{\rm FeK} 
\sim 0.03$ Myr to supply the kinetic energy carried by the OH 
phase, $\mathcal{E}_{\rm OH} \sim 0.13$--$1.5 \times 10^{57}$ erg. Considering the 
amount of wind's mechanical power that ends up stored as thermal energy of the shocked 
ISM or used in the gravitational and expansion work, there is already overlap with 
the OH flow time of 0.1--1 Myr. Moreover, at sub-kpc scales inverse Compton cooling 
by the AGN radiation field can lead to non negligible radiative losses in the shocked 
wind (Ciotti \& Ostriker 1997, 2001; King 2003). With no compelling necessity for 
the AGN to be more powerful than in the recent past, it is then plausible that the \fek 
and OH flows are both part of the same AGN event, related to the ongoing luminous 
stage. Remarkably, this implies that even momentum boosts of a few 
($\mathcal{B}_{\rm OH}$/$\mathcal{B}_{\rm FeK} \sim 2$) can be fully 
compatible with the adiabatic expansion of hot wind bubbles. On the other hand, 
the similarity of the OH and CO outflow velocities might just be serendipitous, 
due to 
the fact that OH is 
driven to slower speeds because it is denser ($f_{\rm OH} \sim 0.01$--0.10 
from Eq.~\ref{e6}), consistent with the hypothesis that some clumps survived the 
previous outflow(s) to be disrupted by the current one. 

Based on Fig.~\ref{wd}, the outflows in \mrk are apparently much easier to explain. This 
relies on two obvious reasons: the SMBH wind is about four times slower, and the outflow 
is extended only out to $\sim$\,3 kpc (Rupke \& Veilleux 2011). The mechanical energy 
of the molecular gas, $\mathcal{E}_{\rm CO} \sim 1.3$--$3 \times 10^{57}$ erg, can be 
released by the AGN in $\mathcal{T}_{\rm Fek}^{\rm CO} \sim 0.1$--1.3 Myr, in perfect agreement 
with the age of the outflow. As the three phases share a common range of kinetic power 
at $\sim$\,0.5--$1\alpha^{-1}$ per cent of $\lagn$ (Table~\ref{st}), \mrk seems, at 
face value, a genuine case of energy conservation. Once a realistic efficiency is 
taken into account, however, the \fek wind is presumably too faint to propel the CO 
outflow, in spite of the lower/upper limit nature of their relative figures. In this 
sense, the BAL component would actually stand out as a more convincing driver, since 
its mass-loss rate can be substantially underestimated if our choice of 
$\mathcal{R}_{\rm BAL}$ is overly conservative. The most critical aspect is thus the 
physical relationship between the \fek and BAL winds. 

An X-ray/UV connection is well-established in AGN for the slower warm absorbers (Crenshaw 
\& Kraemer 2012), where the various features arise in the same, stratified medium, yet \fek 
absorption and BALs are rarely seen together. Consequently, their possible links have been 
never investigated in great detail, also because the initial conditions in numerical 
simulations of AGN outflows do not need to firmly discriminate between the two. While 
evidence for an intrinsic X-ray weakness of BAL quasars is now growing (e.g. Luo et al. 
2014), with \mrk itself fitting into this picture (Teng et al. 2014), the UV spectral 
properties of sources hosting ultra-fast X-ray winds have been largely overlooked so 
far. In \mrk, it would be tempting to speculate that we are witnessing, 
thanks to the X-ray 
weakness, the transition from continuum to line driving, as the ionization in the 
pre-shock SMBH wind drops with distance. This, however, is likely too simplistic once 
compared to the highly complex UV spectrum, which rather points to a clumpy 
outflow (Veilleux et al. 2013a, 2016). In general, \fek and BAL features could even be 
co-spatial at sub-pc scales, if disc winds are strongly inhomogeneous (e.g. Hamann 
et al. 2018).

%%%%%%%%%%%%%%%%%%%%%%%%%%%%%%%%%%%%%%%%%%%%%%%%%%%%%%%%%%%%%%%%%%%%%%%%%%%
\section{Future developments}

In this paper, we have illustrated how the `momentum boost' diagram could become a powerful 
tool not only for attempting a connection between the different manifestations of 
outflows in AGN, but also for reconstructing an empirical history of SMBH accretion. 
Under the hypothesis of a virtually adiabatic expansion of the hot, shocked winds 
(Faucher-Gigu{\`e}re \& Quataert 2012), the proposed application of this diagram is 
very promising, since any major deviations from the expected trends would then be 
directly attributable to fluctuations in the past AGN activity. Our re-examination 
of \iras and \mrk, carried out within a basic analytical framework and intended as 
a preliminary proof of concept, is still incapable of providing any conclusive results, 
owing to the large measure uncertainties and systematics. For these objects, however, 
there is ample room for cutting the error bars down through the acquisition of higher 
quality data, whereas filling the `momentum boost' diagram with other outflow phases 
would allow a better sampling of the velocity and mass outflow rate radial profiles. 
Systematics can instead be overcome only through unbiased campaigns to increase the 
overall statistics, a step that is now widely recognized as mandatory (Cicone et al. 
2018). 

A complementary effort on the theoretical side is also desirable. Future observations 
should be complemented by more detailed simulations to take into account the 
time-evolution of AGN activity, wind mechanical power and ISM properties, with 
specific regard for the peculiar ULIRG environment. Indeed, ULIRGs might easily depart 
from the main assumptions of the current analytical models, such as spherical symmetry, 
and also the appropriateness of the standard SMBH scaling relations is somewhat 
questionable. Even so, ULIRGs remain the best laboratories to study AGN feedback 
in its radiative mode (Fabian et al. 2012). Finding the most powerful outflows in the 
local Universe among ULIRGs (e.g. Gonz{\'a}lez-Alfonso et al. 2017; 
Rupke, G{\"u}ltekin \& Veilleux 2017) is nothing 
but an indirect consequence of cosmic downsizing. It is not our aim here to discuss 
the global effects of AGN outflows, but it is not particularly surprising that their 
impact is rather limited in Seyfert galaxies (Bae et al. 2017; Rosario et al. 2018), 
where $\dot{m}_{\rm A} \sim 0.01$--0.1. In sub-Eddington AGN, continuum scattering is 
not an effective mechanism for driving persistent X-ray winds, which are then unlikely 
to deposit enough energy to sustain a feedback process as required by galaxy evolution 
models. This is reflected by a twofold observational evidence, for which not only are 
\fek winds in Seyferts slower, on average, than in quasars ($\beta < 0.1$ against 
$\beta > 0.2$), but they also have a markedly transient nature. Their detection rate, 
in Seyferts with at least one such claim over multiple observations, is well below 50 
per cent, with just a handful of exceptions (Tombesi et al. 2010; Gofford et al. 
2013). 

Conversely, ultra-fast winds are almost ubiquitous in the high-efficiency accretion 
regime of quasars, and likewise in that of Narrow-Line Seyfert 1 galaxies (Hagino et al. 
2016; Parker et al. 2017) and ultraluminous X-ray sources (Pinto, Middleton \& Fabian 
2016). In ULIRGs, the merger-induced dissipation of ISM angular momentum and the 
subsequent availability of an abundant gas supply for SMBH growth can give rise to 
prolonged and/or recursive Eddington-limited accretion phases. In addition, also 
line driving and radiation pressure on dust (Ishibashi, Fabian \& Maiolino 2018) 
can significantly contribute to push the wind outwards, making it possible in the 
long term to fully erode the dense cold clumps that would otherwise survive owing 
to Rayleigh-Taylor instabilities. Clumps might re-form further out, resulting in an 
inside-out quenching (Tacchella et al. 2015) and even in residual star formation 
(Maiolino et al. 2017). 

The `Eddington connection' between quasars/ULIRGs, Narrow-Line Seyfert 1 galaxies 
(NLS1s) and ultraluminous X-ray sources (ULXs) offers other intriguing hints. In 
some local NLS1s, the models of optical/UV accretion-disc emission overpredict 
the observed X-ray spectrum, suggesting that a large fraction of the accretion power 
is not converted into radiation but employed to launch a wind (Jin et al. 2017). 
Similarly, it has been recently proposed that the so-called ultraluminous supersoft 
sources, arguably the most exotic class of accreting objects, are simply ULXs seen 
through an optically thick wind (Urquhart \& Soria 2016). This sheds new light also 
on the X-ray weakness of quasars, which might be at the same time a requisite for winds 
(for instance, by avoiding over-ionization and so enabling line driving; Murray et al. 
1995) and their consequence. If the broadband spectral energy distribution retains 
the signatures of the blow-out phase, X-ray winds can be studied even when the usual 
\fek tracers are inaccessible. This is of primary importance for the prospect of 
extending the `momentum boost' diagram to the high-redshift AGN populations around 
the peak of the cosmic accretion history, when winds must have shaped the galaxies 
as we see them today. 

These and other relevant aspects of AGN outflows will be investigated in more 
detail from both an observational and a theoretical standpoint in future works.

%%%%%%%%%%%%%%%%%%%%%%%%%%%%%%%%%%%%%%%%%%%%%%%%%%%%%%%%%%%%%%%%%%%%%%%%%%%
\section*{Acknowledgments}
We thank the anonymous referee for the insightful and constructive comments that 
helped us improve the clarity of the paper, and Chiara Feruglio, Nozomu Kawakatu, 
Eleonora Sani, Francesco Tombesi and XianZhong Zheng for sharing information or useful 
discussion. EN acknowledges funding from the European Union's Horizon 2020 research and 
innovation programme under the Marie Sk\l{}odowska-Curie grant agreement no. 664931, and 
is grateful to the Centre for Extragalactic Astronomy, Durham University, for recurrent 
hospitality. KZ is funded by the Research Council Lithuania through the grant no. MIP-17-78.

%%%%%%%%%%%%%%%%%%%%%%%%%%%%%%%%%%%%%%%%%%%%%%%%%%%%%%%%%%%%%%%%%%%%%%%%%%%

%%%%%%%%%%%%%%%%%%%%%%%%%%%%%%%%%%%%%%%%%%%%%%%%%%%%%%%%%%%%%%%%%%%%%%%%%%%

\label{lastpage}

%%%%%%%%%%%%%%%%%%%%%%%%%%%%%%%%%%%%%%%%%%%%%%%%%%%%%%%%%%%%%%%%%%%%%%%%%%%

\begin{thebibliography}{}
\bibitem[\protect\citeauthoryear{Aalto et al.}{2015}]{2015A&A...574A..85A} 
Aalto S., et al., 2015, A\&A, 574, A85 
\bibitem[\protect\citeauthoryear{Bentz et al.}{2013}]{2013ApJ...767..149B} 
Bentz M.~C., et al., 2013, ApJ, 767, 149 
\bibitem[\protect\citeauthoryear{Bae et al.}{2017}]{2017ApJ...837...91B} 
Bae H.-J., Woo J.-H., Karouzos M., Gallo E., Flohic H., Shen Y., Yoon S.-J., 2017, ApJ, 837, 91 
\bibitem[\protect\citeauthoryear{Bischetti et al.}{2017}]{2017A&A...598A.122B} 
Bischetti M., et al., 2017, A\&A, 598, A122 
\bibitem[\protect\citeauthoryear{Capelo et al.}{2015}]{2015MNRAS.447.2123C} 
Capelo P.~R., Volonteri M., Dotti M., Bellovary J.~M., Mayer L., Governato F., 2015, MNRAS, 447, 2123 
\bibitem[\protect\citeauthoryear{Carniani et al.}{2015}]{2015A&A...580A.102C} 
Carniani S., et al., 2015, A\&A, 580, A102
%\bibitem[\protect\citeauthoryear{Chartas et al.}{2002}]{2002ApJ...579..169C} 
%Chartas G., Brandt W.~N., Gallagher S.~C., Garmire G.~P., 2002, ApJ, 579, 169 
%\bibitem[\protect\citeauthoryear{Chartas et al.}{2009}]{2009ApJ...706..644C} 
%Chartas G., Saez C., Brandt W.~N., Giustini M., Garmire G.~P., 2009, ApJ, 706, 644 
\bibitem[\protect\citeauthoryear{Cicone et al.}{2014}]{2014A&A...562A..21C} 
Cicone C., et al., 2014, A\&A, 562, A21 
\bibitem[\protect\citeauthoryear{Cicone et al.}{2018}]{2018NatAs...2..176C} 
Cicone C., Brusa M., Ramos Almeida C., Cresci G., Husemann B., Mainieri V., 2018, NatAs, 2, 176
\bibitem[\protect\citeauthoryear{Ciotti \& Ostriker}{1997}]{1997ApJ...487L.105C} 
Ciotti L., Ostriker J.~P., 1997, ApJ, 487, L105
\bibitem[\protect\citeauthoryear{Ciotti \& Ostriker}{2001}]{2001ApJ...551..131C} 
Ciotti L., Ostriker J.~P., 2001, ApJ, 551, 131 
\bibitem[\protect\citeauthoryear{Crenshaw \& Kraemer}{2012}]{2012ApJ...753...75C} 
Crenshaw D.~M., Kraemer S.~B., 2012, ApJ, 753, 75 
\bibitem[\protect\citeauthoryear{Dasyra et al.}{2006}]{2006ApJ...651..835D} 
Dasyra K.~M., et al., 2006, ApJ, 651, 835
\bibitem[\protect\citeauthoryear{Downes \& Solomon}{1998}]{1998ApJ...507..615D}
Downes D., Solomon P.~M., 1998, ApJ, 507, 615 
\bibitem[\protect\citeauthoryear{Dunn et al.}{2010}]{2010ApJ...709..611D} 
Dunn J.~P., et al., 2010, ApJ, 709, 611 
\bibitem[\protect\citeauthoryear{Fabian}{2012}]{2012ARA&A..50..455F} 
Fabian A.~C., 2012, ARA\&A, 50, 455 
\bibitem[\protect\citeauthoryear{Faucher-Gigu{\`e}re \& Quataert}{2012}]{2012MNRAS.425..605F} 
Faucher-Gigu{\`e}re C.-A., Quataert E., 2012, MNRAS, 425, 605 
\bibitem[\protect\citeauthoryear{Feruglio et al.}{2010}]{2010A&A...518L.155F} 
Feruglio C., Maiolino R., Piconcelli E., Menci N., Aussel H., Lamastra A., Fiore F., 2010, A\&A, 518, L155 
\bibitem[\protect\citeauthoryear{Feruglio et al.}{2015}]{2015A&A...583A..99F} 
Feruglio C., et al., 2015, A\&A, 583, A99 
%\bibitem[\protect\citeauthoryear{Feruglio et al.}{2017}]{2017A&A...608A..30F} 
%Feruglio C., et al., 2017, A\&A, 608, A30 
\bibitem[\protect\citeauthoryear{Fiore et al.}{2017}]{2017A&A...601A.143F} 
Fiore F., et al., 2017, A\&A, 601, A143 
\bibitem[\protect\citeauthoryear{Fischer et al.}{2010}]{2010A&A...518L..41F} 
Fischer J., et al., 2010, A\&A, 518, L41 
\bibitem[\protect\citeauthoryear{Gabor \& Bournaud}{2014}]{2014MNRAS.441.1615G} 
Gabor J.~M., Bournaud F., 2014, MNRAS, 441, 1615 
%\bibitem[\protect\citeauthoryear{Gaspari, Ruszkowski, \& Oh}{2013}]{2013MNRAS.432.3401G} 
%Gaspari M., Ruszkowski M., Oh S.~P., 2013, MNRAS, 432, 3401 
\bibitem[\protect\citeauthoryear{Gibson et al.}{2009}]{2009ApJ...692..758G} 
Gibson R.~R., et al., 2009, ApJ, 692, 758 
\bibitem[\protect\citeauthoryear{Gofford et al.}{2013}]{2013MNRAS.430...60G} 
Gofford J., Reeves J.~N., Tombesi F., Braito V., Turner T.~J., Miller L., Cappi M., 2013, MNRAS, 430, 60
%\bibitem[\protect\citeauthoryear{Gofford et al.}{2015}]{2015MNRAS.451.4169G} 
%Gofford J., Reeves J.~N., McLaughlin D.~E., Braito V., Turner T.~J., Tombesi F., Cappi M., 2015, MNRAS, 451, 4169 
\bibitem[\protect\citeauthoryear{Gonz{\'a}lez-Alfonso et al.}{2014}]{2014A&A...561A..27G} 
Gonz{\'a}lez-Alfonso E., et al., 2014, A\&A, 561, A27 
\bibitem[\protect\citeauthoryear{Gonz{\'a}lez-Alfonso et al.}{2017}]{2017ApJ...836...11G} 
Gonz{\'a}lez-Alfonso E., et al., 2017, ApJ, 836, 11 
%\bibitem[\protect\citeauthoryear{Hagino et al.}{2017}]{2017MNRAS.468.1442H} 
%Hagino K., Done C., Odaka H., Watanabe S., Takahashi T., 2017, MNRAS, 468, 1442 
\bibitem[\protect\citeauthoryear{Hamann et al.}{2018}]{2018MNRAS.tmp...33H} 
Hamann F., Chartas G., Reeves J., Nardini E., 2018, MNRAS, in press
\bibitem[\protect\citeauthoryear{Hao et al.}{2005}]{2005ApJ...625...78H} 
Hao C.~N., Xia X.~Y., Mao S., Wu H., Deng Z.~G., 2005, ApJ, 625, 78 
\bibitem[\protect\citeauthoryear{Hagino et al.}{2016}]{2016MNRAS.461.3954H} 
Hagino K., Odaka H., Done C., Tomaru R., Watanabe S., Takahashi T., 2016, MNRAS, 461, 3954 
\bibitem[\protect\citeauthoryear{Harrison}{2017}]{2017NatAs...1E.165H} 
Harrison C.~M., 2017, NatAs, 1, 0165 
\bibitem[\protect\citeauthoryear{Harrison et al.}{2014}]{2014MNRAS.441.3306H} 
Harrison C.~M., Alexander D.~M., Mullaney J.~R., Swinbank A.~M., 2014, MNRAS, 441, 3306 
\bibitem[\protect\citeauthoryear{Harrison et al.}{2018}]{2018NatAs...2..198H} 
Harrison C.~M., Costa T., Tadhunter C.~N., Fl{\"u}tsch A., Kakkad D., Perna M., Vietri G., 2018, NatAs, 2, 198
%\bibitem[\protect\citeauthoryear{Kallman \& Bautista}{2001}]{2001ApJS..133..221K} 
%Kallman T., Bautista M., 2001, ApJS, 133, 221 
\bibitem[\protect\citeauthoryear{Ishibashi, Fabian, \& Maiolino}{2018}]{2018MNRAS.476..512I} 
Ishibashi W., Fabian A.~C., Maiolino R., 2018, MNRAS, 476, 512 
\bibitem[\protect\citeauthoryear{Jin et al.}{2017}]{2017MNRAS.471..706J} 
Jin C., Done C., Ward M., Gardner E., 2017, MNRAS, 471, 706
\bibitem[\protect\citeauthoryear{Kawakatu, Imanishi, \& Nagao}{2007}]{2007ApJ...661..660K} 
Kawakatu N., Imanishi M., Nagao T., 2007, ApJ, 661, 660
\bibitem[\protect\citeauthoryear{King}{2003}]{2003ApJ...596L..27K} 
King A., 2003, ApJ, 596, L27 
\bibitem[\protect\citeauthoryear{King}{2010}]{2010MNRAS.402.1516K} 
King A.~R., 2010, MNRAS, 402, 1516 
\bibitem[\protect\citeauthoryear{King \& Nixon}{2015}]{2015MNRAS.453L..46K} 
King A., Nixon C., 2015, MNRAS, 453, L46 
%\bibitem[\protect\citeauthoryear{King \& Pounds}{2003}]{2003MNRAS.345..657K} 
%King A.~R., Pounds K.~A., 2003, MNRAS, 345, 657 
\bibitem[\protect\citeauthoryear{King \& Pounds}{2015}]{2015ARA&A..53..115K} 
King A., Pounds K., 2015, ARA\&A, 53, 115 
\bibitem[\protect\citeauthoryear{King, Zubovas, \& Power}{2011}]{2011MNRAS.415L...6K} 
King A.~R., Zubovas K., Power C., 2011, MNRAS, 415, L6 
\bibitem[\protect\citeauthoryear{Kormendy \& Ho}{2013}]{2013ARA&A..51..511K} 
Kormendy J., Ho L.~C., 2013, ARA\&A, 51, 511 
\bibitem[\protect\citeauthoryear{Krongold et al.}{2007}]{2007ApJ...659.1022K} 
Krongold Y., Nicastro F., Elvis M., Brickhouse N., Binette L., Mathur S., Jim{\'e}nez-Bail{\'o}n E., 2007, ApJ, 659, 1022 
\bibitem[\protect\citeauthoryear{Landt et al.}{2013}]{2013MNRAS.432..113L} 
Landt H., Ward M.~J., Peterson B.~M., Bentz M.~C., Elvis M., Korista K.~T., Karovska M., 2013, MNRAS, 432, 113
\bibitem[\protect\citeauthoryear{Leighly et al.}{2014}]{2014ApJ...788..123L} 
Leighly K.~M., Terndrup D.~M., Baron E., Lucy A.~B., Dietrich M., Gallagher S.~C., 2014, ApJ, 788, 123 
\bibitem[\protect\citeauthoryear{L{\'{\i}}pari et al.}{2003}]{2003MNRAS.340..289L} 
L{\'{\i}}pari S., Terlevich R., D{\'{\i}}az R.~J., Taniguchi Y., Zheng W., Tsvetanov Z., Carranza G., Dottori H., 2003, MNRAS, 340, 289
\bibitem[\protect\citeauthoryear{Lipari et al.}{2009}]{2009MNRAS.392.1295L} 
Lipari S., et al., 2009, MNRAS, 392, 1295 
\bibitem[\protect\citeauthoryear{Liu et al.}{2013}]{2013MNRAS.436.2576L} 
Liu G., Zakamska N.~L., Greene J.~E., Nesvadba N.~P.~H., Liu X., 2013, MNRAS, 436, 2576 
\bibitem[\protect\citeauthoryear{Luo et al.}{2014}]{2014ApJ...794...70L} 
Luo B., et al., 2014, ApJ, 794, 70 
\bibitem[\protect\citeauthoryear{Maiolino et al.}{2012}]{2012MNRAS.425L..66M} 
Maiolino R., et al., 2012, MNRAS, 425, L66 
\bibitem[\protect\citeauthoryear{Maiolino et al.}{2017}]{2017Natur.544..202M} 
Maiolino R., et al., 2017, Natur, 544, 202 
\bibitem[\protect\citeauthoryear{Morganti et al.}{2016}]{2016A&A...593A..30M} 
Morganti R., Veilleux S., Oosterloo T., Teng S.~H., Rupke D., 2016, A\&A, 593, A30 
\bibitem[\protect\citeauthoryear{Murray et al.}{1995}]{1995ApJ...451..498M} 
Murray N., Chiang J., Grossman S.~A., Voit G.~M., 1995, ApJ, 451, 498 
\bibitem[\protect\citeauthoryear{Nardini et al.}{2010}]{2010MNRAS.405.2505N} 
Nardini E., Risaliti G., Watabe Y., Salvati M., Sani E., 2010, MNRAS, 405, 2505
\bibitem[\protect\citeauthoryear{Nardini et al.}{2015}]{2015Sci...347..860N} 
Nardini E., et al., 2015, Sci, 347, 860 
\bibitem[\protect\citeauthoryear{Nayakshin, Power, \& King}{2012}]{2012ApJ...753...15N} 
Nayakshin S., Power C., King A.~R., 2012, ApJ, 753, 15 
\bibitem[\protect\citeauthoryear{Parker et al.}{2017}]{2017Natur.543...83P} 
Parker M.~L., et al., 2017, Natur, 543, 83 
\bibitem[\protect\citeauthoryear{Pinto, Middleton, \& Fabian}{2016}]{2016Natur.533...64P} 
Pinto C., Middleton M.~J., Fabian A.~C., 2016, Natur, 533, 64 
\bibitem[\protect\citeauthoryear{Reynolds}{2012}]{2012ApJ...759L..15R} 
Reynolds C.~S., 2012, ApJ, 759, L15 
\bibitem[\protect\citeauthoryear{Richings \& Faucher-Gigu{\`e}re}{2017}]{2017arXiv171009433R} 
Richings A.~J., Faucher-Gigu{\`e}re C.-A., 2017, arXiv, arXiv:1710.09433 
\bibitem[\protect\citeauthoryear{Rosario et al.}{2018}]{2018MNRAS.473.5658R} 
Rosario D.~J., et al., 2018, MNRAS, 473, 5658 
\bibitem[\protect\citeauthoryear{Rothberg et al.}{2013}]{2013ApJ...767...72R} 
Rothberg B., Fischer J., Rodrigues M., Sanders D.~B., 2013, ApJ, 767, 72
\bibitem[\protect\citeauthoryear{Rupke \& Veilleux}{2011}]{2011ApJ...729L..27R} 
Rupke D.~S.~N., Veilleux S., 2011, ApJ, 729, L27 
\bibitem[\protect\citeauthoryear{Rupke \& Veilleux}{2013}]{2013ApJ...768...75R} 
Rupke D.~S.~N., Veilleux S., 2013, ApJ, 768, 75
\bibitem[\protect\citeauthoryear{Rupke, G{\"u}ltekin, \& Veilleux}{2017}]{2017ApJ...850...40R} 
Rupke D.~S.~N., G{\"u}ltekin K., Veilleux S., 2017, ApJ, 850, 40 
\bibitem[\protect\citeauthoryear{Rupke, Veilleux, \& Sanders}{2005}]{2005ApJ...632..751R} 
Rupke D.~S., Veilleux S., Sanders D.~B., 2005, ApJ, 632, 751
\bibitem[\protect\citeauthoryear{Schawinski et al.}{2015}]{2015MNRAS.451.2517S} 
Schawinski K., Koss M., Berney S., Sartori L.~F., 2015, MNRAS, 451, 2517 
\bibitem[\protect\citeauthoryear{Spoon \& Holt}{2009}]{2009ApJ...702L..42S} 
Spoon H.~W.~W., Holt J., 2009, ApJ, 702, L42 
\bibitem[\protect\citeauthoryear{Sturm et al.}{2011}]{2011ApJ...733L..16S} 
Sturm E., et al., 2011, ApJ, 733, L16 
\bibitem[\protect\citeauthoryear{Tacchella et al.}{2015}]{2015Sci...348..314T} 
Tacchella S., et al., 2015, Sci, 348, 314 
\bibitem[\protect\citeauthoryear{Takeuchi, Ohsuga, \& Mineshige}{2013}]{2013PASJ...65...88T} 
Takeuchi S., Ohsuga K., Mineshige S., 2013, PASJ, 65, 88
\bibitem[\protect\citeauthoryear{Teng et al.}{2014}]{2014ApJ...785...19T} 
Teng S.~H., et al., 2014, ApJ, 785, 19 
\bibitem[\protect\citeauthoryear{Tombesi et al.}{2010}]{2010A&A...521A..57T} 
Tombesi F., Cappi M., Reeves J.~N., Palumbo G.~G.~C., Yaqoob T., Braito V., Dadina M., 2010, A\&A, 521, A57 
\bibitem[\protect\citeauthoryear{Tombesi et al.}{2015}]{2015Natur.519..436T} 
Tombesi F., Mel{\'e}ndez M., Veilleux S., Reeves J.~N., Gonz{\'a}lez-Alfonso E., Reynolds C.~S., 2015, Natur, 519, 436
\bibitem[\protect\citeauthoryear{Tombesi et al.}{2017}]{2017ApJ...850..151T} Tombesi F., Veilleux S., Mel{\'e}ndez M., Lohfink A., Reeves J.~N., Piconcelli E., Fiore F., Feruglio C., 2017, ApJ, 850, 151
\bibitem[\protect\citeauthoryear{Urquhart \& Soria}{2016}]{2016MNRAS.456.1859U} 
Urquhart R., Soria R., 2016, MNRAS, 456, 1859 
\bibitem[\protect\citeauthoryear{Veilleux et al.}{2009}]{2009ApJ...701..587V} 
Veilleux S., et al., 2009a, ApJ, 701, 587
\bibitem[\protect\citeauthoryear{Veilleux et al.}{2009}]{2009ApJS..182..628V} 
Veilleux S., et al., 2009b, ApJS, 182, 628--666 
\bibitem[\protect\citeauthoryear{Veilleux et al.}{2013}]{2013ApJ...764...15V} 
Veilleux S., et al., 2013a, ApJ, 764, 15
\bibitem[\protect\citeauthoryear{Veilleux et al.}{2013}]{2013ApJ...776...27V} 
Veilleux S., et al., 2013b, ApJ, 776, 27 
\bibitem[\protect\citeauthoryear{Veilleux et al.}{2016}]{2016ApJ...825...42V} 
Veilleux S., Mel{\'e}ndez M., Tripp T.~M., Hamann F., Rupke D.~S.~N., 2016, ApJ, 825, 42 
\bibitem[\protect\citeauthoryear{Veilleux et al.}{2017}]{2017ApJ...843...18V} 
Veilleux S., Bolatto A., Tombesi F., Mel{\'e}ndez M., Sturm E., Gonz{\'a}lez-Alfonso E., Fischer J., Rupke D.~S.~N., 2017, ApJ, 843, 18 
\bibitem[\protect\citeauthoryear{Zheng et al.}{2002}]{2002AJ....124...18Z} 
Zheng X.~Z., Xia X.~Y., Mao S., Wu H., Deng Z.~G., 2002, AJ, 124, 18 
\bibitem[\protect\citeauthoryear{Zubovas}{2018}]{2018MNRAS.473.3525Z} 
Zubovas K., 2018, MNRAS, 473, 3525 
\bibitem[\protect\citeauthoryear{Zubovas \& King}{2012}]{2012ApJ...745L..34Z} 
Zubovas K., King A., 2012, ApJ, 745, L34
\bibitem[\protect\citeauthoryear{Zubovas \& King}{2014}]{2014MNRAS.439..400Z} 
Zubovas K., King A.~R., 2014, MNRAS, 439, 400 
\bibitem[\protect\citeauthoryear{Zubovas \& King}{2016}]{2016MNRAS.462.4055Z} 
Zubovas K., King A., 2016, MNRAS, 462, 4055 
\end{thebibliography}
\end{document}